\documentclass[12pt]{l4dc2021}

\title[Machine learning based compensation of static effects of residual dynamics]{Learning-based feedforward augmentation for steady state rejection of residual dynamics on a nanometer-accurate planar actuator system}
\usepackage{times}
\usepackage{amsfonts}
\usepackage{graphicx}
\usepackage{graphics}

\makeatletter
 \let\Ginclude@graphics\@org@Ginclude@graphics 
\makeatother

\makeatletter
\let\@fnsymbol\@arabic
\makeatother

\definecolor{thoughtProvYellow}{rgb}{0.9255    0.8157    0.4706}
\definecolor{thoughtProvOrange}{rgb}{0.8510    0.3569    0.2627}
\definecolor{thoughtProvRed}{rgb}{0.7529    0.1608    0.2588}
\definecolor{thoughtProvDark}{rgb}{0.3294    0.1412    0.2157}
\definecolor{thoughtProvGreen}{rgb}{0.3255    0.4667    0.4784}

\definecolor{newPaletteBlue}{rgb}{0.3020 0.5220    0.7410}
\definecolor{newPaletteOrange}{rgb}{0.796 0.388 0.094}
\definecolor{newPaletteGreen}{rgb}{0.486 0.667 0.176} 
\definecolor{newPaletteYellow}{rgb}{0.9610 0.8900 0.3370}
\definecolor{thoughtProvPink}{rgb}{0.9492 0.0.800 0.7734}
\definecolor{thoughtProvYellowLight}{rgb}{0.9727 0.941 0.8320}

\newcommand{\myRule}[1]{\textcolor{#1}{\rule[1pt]{4mm}{0.8mm}}}

\newcommand\figref{Figure~\ref}

 \author{\Name{Ioannis Proimadis}\thanks{VDL ETG Technology $\&$ Development, Eindhoven 5651GH, The Netherlands } \Email{jproimadis@gmail.com}\\ 
\Name{Yorick Broens}\thanks{Control Systems Group, Eindhoven University of Technology, 5600MB, The Netherlands} \Email{Y.L.C.Broens@tue.nl} 
\\
\Name{Roland T\'oth}\textcolor{blue}{\footnotemark[2]}\textsuperscript{,}\thanks{Systems and Control Laboratory, Institute for Computer Science and Control, Kende u. 13-17, H-1111 Budapest, Hungary} \Email{R.Toth@tue.nl} \\
\Name{Hans Butler}\textcolor{blue}{\footnotemark[2]}\textsuperscript{,}\thanks{ASML, Veldhoven 5504DR, The Netherlands} \Email{H.Butler@tue.nl}}

\begin{document}
\maketitle
\vspace*{-6.5mm}
\begin{abstract}
Growing demands in the semiconductor industry result in the need for enhanced performance of lithographic equipment. However, position tracking accuracy of high precision mechatronics is often limited by the presence of disturbance sources, which originate from unmodelled or unforeseen deterministic environmental effects.
To negate the effects of these disturbances, a learning based feedforward controller is employed, where the underlying control policy is estimated from experimental data based on Gaussian Process regression. The proposed approach exploits the property of including prior knowledge on the expected steady state behaviour of residual dynamics in terms of kernel selection. Corresponding hyper-parameters are optimized using the maximization of the marginalized likelihood. Consequently, the learned function is employed as augmentation of the currently employed rigid body feedforward controller. The effectiveness of the augmentation is experimentally validated on a magnetically levitated planar motor stage.
The results of this paper demonstrate the benefits and possibilities of machine-learning based approaches for compensation of static effects, which originate from residual dynamics, such that  position tracking performance for moving-magnet planar motor actuators is improved.
\label{Abstract}
\end{abstract}

\vspace*{-4mm}
\section{Introduction}
\vspace*{-1.5mm}
\hspace*{2mm} In high-precision lithography, production of integrated circuits is realized by projecting extreme ultraviolet light on a silicon wafer using projection optics. In order to achieve high throughput and high reliability, 
the silicon wafer is positioned under the projection optics using the wafer stage module, which is a planar motor system  that is capable of achieving nanometer accuracy in position tracking. The growing demands in the semiconductor industry result in the necessity to increase throughput, while still maintaining accurate positioning. In order to meet the throughput demands, highly aggressive acceleration profiles are required, which introduce high-frequent position tracking errors due to the limited stiffness of the mechanical structure. 
The currently implemented state-of-the-art planar stage configuration, which is based on the design proposed by \cite{370Cho} and further discussed by \cite{172Compter}, relies on a double stroke mechanism, where a magnetically levitated moving-coil motor is used for the coarse positioning of the mover. For fine positioning of the mover, the short stroke motor is used, which is actuated by voice coils.
\\
\hspace*{2mm} As an alternative, planar motors based on a moving-magnet configuration have been investigated in recent years, see \cite{344Boeij}. In contrast with the moving-coil configuration, the moving-magnet configuration is comprised of a stator base and a freely floating magnet plate. The absence of physical connections between the mover and the environment results in a significant reduction of induced disturbances. Additionally, the moving-magnet configuration allows for a smaller and lighter moving body, see \cite{Proimadis-phd} and \cite{Rovers-phd}, such that high accelerations can be achieved with relatively low power demands.  \\ 
\hspace*{2mm} However, these advantages come at the cost of introducing additional complexity.
Magnetically levitated planar motors exhibit complex non-linear multi-physical effects and are subject to various disturbances, which are machine specific in terms of the design of the magnet and coil arrays, see \cite{Rovers-phd}. The highly complex dynamics can only be approximately modelled based on first principle knowledge and therefore the  achievable position tracking performance is limited.
\\
\hspace*{2mm} In order to achieve nanometer position tracking in magnetically levitated planar motor systems, feedforward control plays a crucial role, see \cite{207Clayton}. However, for accurate design of a feedforward controller, an accurate plant model is required, which, for moving-magnet planar actuator systems, is not trivial to obtain due to the complexity of the moving-magnet configuration. Therefore, standard feedforward control strategies do not provide the desired position tracking performance due to model mismatch between the first-principle based model and the real system.
\\
\hspace*{2mm} In order to improve position tracking performance, learning based strategies can be employed to construct a feedforward policy, see \cite{mooren2020gaussian}, \cite{inproceedings} and \cite{Proimadis-phd}. By such a learning-based approach, the steady state behaviour of residual dynamics is captured by viewing it as a load disturbance. Then this load disturbance can be modelled using the Gaussian Process (GP) framework,  which is advantageous since it ensures uncertainty bounds, such that reliability of the GP model is guaranteed.
 \\
\hspace*{2mm} Capturing the behaviour of unforeseen dynamics as a function of generalized coordinates allows for augmentation of the currently employed rigid body feedforward controller, such that static effects of residual dynamics of the magnetically levitated planar motor system are compensated for. 
\\
\hspace*{2mm} This paper is organized as follows. In Section \ref{Section2}, a brief description of the magnetically levitated planar motor system is presented. Section \ref{Section3} presents 
the GP-based modelling of steady state behaviour of residual dynamics. Section \ref{ch:ExperimentalResults} describes the experimental validation of the designed feedforward augmentation, where the experiments are performed on a magnetically levitated planar actuator. Lastly, in Section \ref{ch:Conclusions} the conclusions are drawn.

\label{Introduction}

\vspace*{-2.5mm}
\section{Magnetically levitated planar motor system}
\label{Section2}
\vspace*{-1.5mm}
\subsection{System overview}
\begin{figure}[h]
\begin{center}
\includegraphics[width=0.95\linewidth]{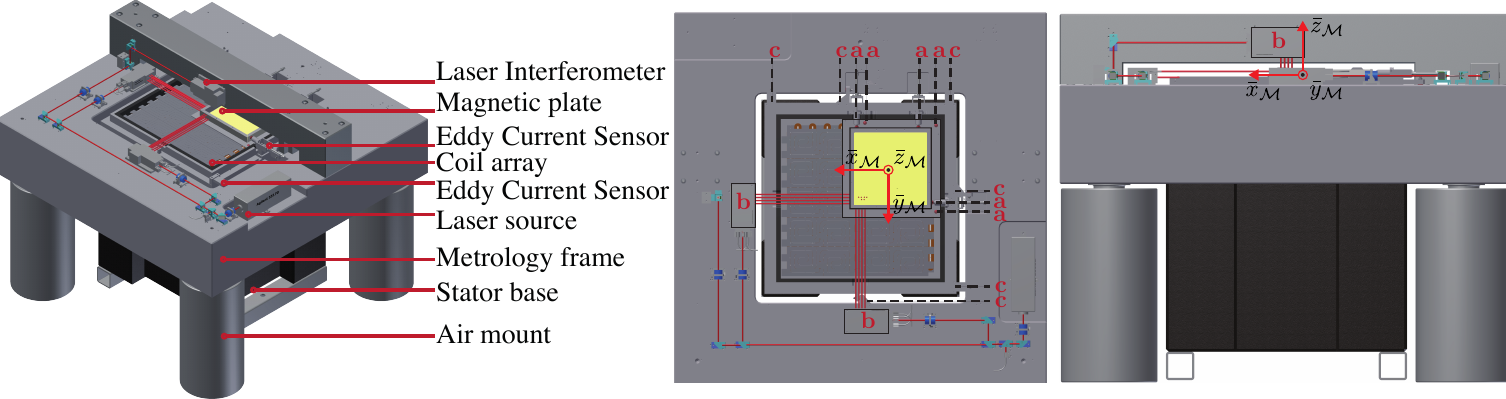}
\end{center}
\caption{Schematic representation of the NAPAS setup, with a) ECSs used for initialization  b) LIFM devices . c) ECS used to capture displacements between $\mathcal{C}-\mathcal{M}$. $\bar{x}_\mathcal{M}$, $\bar{y}_\mathcal{M}$ and $\bar{z}_\mathcal{M}$ denote the metrology coordinate frame.}
\label{fig:NAPAS}
\end{figure}

\vspace*{-1.5mm}
\hspace*{2mm} The Nanometer Accurate Planar Actuator System (NAPAS) prototype, which is based on a moving-magnet configuration, is illustrated in Figure \ref{fig:NAPAS}. The NAPAS prototype consists of three separate coordinate frames: The stator base ($\mathcal{C}$), the translator ($\mathcal{T}$) and the metrology frame ($\mathcal{M}$). The stator base is a double layer coil array, consisting of 160 coils of which 40 coils are simultaneously activated at every time instant using 40 power converters, depending on the relative position of the translator. Proper actuation of the coils offers the means of both stabilization and propulsion of the magnet plate in 6 Degrees of Freedom (DOFs). The metrology frame, which rests on air mounts to suppress effects of any floor induced disturbances, is used as a global reference frame, such that position tracking accuracy can be evaluated. On the metrology frame, 9 laser interferometers (LIFMs) are mounted to measure the relative displacement of the translator with respect to the metrology frame. Additionally, two sets of eddy current sensors (ECS) are mounted as auxiliary measurement systems, where one set of ECS is used for initialization of the NAPAS prototype
and the second set of ECS is used to capture the displacements between the stator base and the metrology frame ($\mathcal{C}$-$\mathcal{M}$). In order to relate the coordinate frames to each other, rigid body coordinate frame transformations are applied, see \cite{Murray-book}.

\vspace*{-5mm}

\noindent
\hspace*{2mm} The dynamic behaviour of the planar motor is governed by both electromagnetic and mechanical phenomena. The electromagnetic interaction describes the relation between the input currents in the coils and the resulting magnet force distribution on the magnet plate of the translator. Moreover, the mechanical model describes the relation between the aforementioned force distribution and the resulting motion of the translator.
\vspace*{-3mm}
\subsection{Electromagnetic interaction}
\vspace*{-1.5mm}
\hspace*{2mm} The actuation of the electromagnetically-levitated planar motor is realized by 40 power amplifiers, which independently control the current supplied to the corresponding 40 active coils. The relation between the supplied currents, $\boldsymbol{i} \in \mathbb{R}^{40}$, and the resulting force vector, is described by the Lorentz force principle \citep{173Rovers}. Under the rigid body assumption, the force distribution, exerted on the magnet plate, is equivalently described as a force and torque vector around the center of mass \citep{Murray-book}. Since the position of the plate is evaluated with respect to the metrology frame, $\mathcal{M}$, it is convenient to define the force/torque vector with respect to the same frame.  This force/torque vector is defined as
\vspace*{-2mm}
\begin{align}
\boldsymbol{W}_{\mathcal{M}} = \begin{bmatrix}
F_x & F_y & F_z & \tau_x & \tau_y & \tau_z
\end{bmatrix}^\top,
\end{align}
where $F_x$ and $\tau_x$ are the force and torque acting on the $x$ axis (similarly for the other axes), expressed in the metrology frame. Then, the electromagnetic interaction is given by
\begin{equation}
\begin{split}
\boldsymbol{W} _{\mathcal{M}}
&= 
R^{\mathcal{M}}_{\mathcal{C}} 
M^{\mathrm{r}} \left( \boldsymbol{q}^{\mathcal{C}}_{\mathcal{T}} \right) \boldsymbol{i},
\end{split}
\label{eq:rigidbodyEM}
\end{equation}

\noindent
where $M^{\mathrm{r}}: \mathbb{R}^6 \mapsto \mathbb{R}^{6 \times 40}$ is a position dependent matrix, which depends on the relative position between the stator base and the mover, which is denoted as $\boldsymbol{q}^{\mathcal{C}}_{\mathcal{T}}$ = $\begin{bmatrix} x_{\mathcal{M}}^{\mathcal{C}}
&
y_{\mathcal{M}}^{\mathcal{C}}
&
z_{\mathcal{M}}^{\mathcal{C}}
&
\chi_{\mathcal{M}}^{\mathcal{C}}
&
\psi_{\mathcal{M}}^{\mathcal{C}}
&
\zeta_{\mathcal{M}}^{\mathcal{C}}
\end{bmatrix}^{\top}$. Moreover, since the force/torque vector is expressed with respect to the coil frame, the rotation matrix $R^{\mathcal{M}}_{\mathcal{C}}$ transforms the force and torque vector to the metrology coordinate frame.
\vspace*{-3mm}
\subsection{Mechanical system}
\vspace*{-1.5mm}
\hspace*{2mm} The kinematics of the translator are derived using Euler-Lagrange modelling strategies, see \cite{184Jeltsema}. To linearise the dynamics of the translator, a small angle approximation is applied, since in lithographic equipment, rotational axes are actively steered to zero. Consequently, rotation angles are neglectable during operation, see \cite{176Butler}. 
The resulting state space representation is expressed by  (\ref{eq:SSrepresentationDynamics}), where $\boldsymbol{q}^{\mathcal{M}}_{\mathcal{T}}$ = $\begin{bmatrix} x_{\mathcal{T}}^{\mathcal{M}}
&
y_{\mathcal{T}}^{\mathcal{M}}
&
z_{\mathcal{T}}^{\mathcal{M}}
&
\chi_{\mathcal{T}}^{\mathcal{M}}
&
\psi_{\mathcal{T}}^{\mathcal{M}}
&
\zeta_{\mathcal{T}}^{\mathcal{M}}
\end{bmatrix}^{\top}$ denotes the relative position of the mover with respect to the metrology frame and $\boldsymbol{G}_\mathcal{M}$ denotes the gravity compensation vector. 

\begin{equation}
\begin{bmatrix}
\dot{\boldsymbol{q}}_\mathcal{T}^\mathcal{M} \\ 
\ddot{\boldsymbol{q}}_\mathcal{T}^\mathcal{M}
\end{bmatrix} = 
\begin{bmatrix}
0 & I \\ 0 & 0
\end{bmatrix}
\begin{bmatrix}
\boldsymbol{q}_\mathcal{T}^\mathcal{M} \\ \dot{\boldsymbol{q}}_\mathcal{T}^\mathcal{M}
\end{bmatrix} + \begin{bmatrix}0 \\ B \end{bmatrix}
(\boldsymbol{W}_\mathcal{M}-\boldsymbol{G}_\mathcal{M})
\label{eq:SSrepresentationDynamics}
\end{equation}

\subsection{Control Principles}
\begin{figure}[h]
\begin{center}
\includegraphics[width=0.8\linewidth]{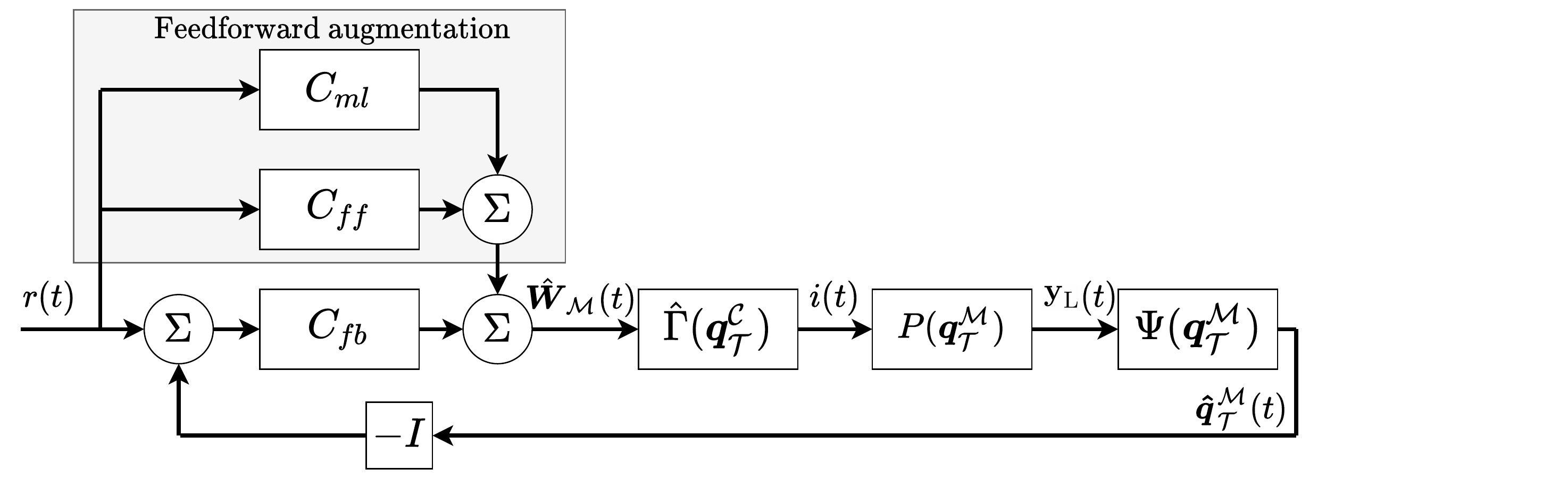}
\end{center}
\vspace*{-5mm}
\caption{Schematic representation of the control architecture of the NAPAS prototype.}
\label{fig:CtrlLoop}
\end{figure}

\vspace*{-1.5mm}
\hspace*{2mm} The control architecture of the NAPAS prototype, which is based on the standard motion control design for wafer scanners, is depicted in Figure \ref{fig:CtrlLoop}. In order to apply SISO control strategies for first-principles feedback control design, the system is rigid body decoupled (\cite{190Steinbuch}), where the commutation $\hat{\Gamma}(\boldsymbol{q}^{\mathcal{C}}_{\mathcal{T}}) = \begin{bmatrix}R^{\mathcal{M}}_{\mathcal{C}} 
M^{\mathrm{r}} ( \boldsymbol{q}^{\mathcal{C}}_{\mathcal{T}} ) \end{bmatrix}^\dagger$ denotes the actuator decoupling (mapping from control forces to required currents, $\hat{\boldsymbol{W}}_{\mathcal{M}}\in \mathbb{R}^6 \mapsto \boldsymbol{i} \in \mathbb{R}^{40}$) and the state reconstruction $\Psi(\boldsymbol{q}_\mathcal{T}^\mathcal{M})$ denotes the sensor decoupling (mapping from independent LIFM measurements to the estimated physical axes, $\boldsymbol{\mathrm{y}}_\mathrm{L} \in \mathbb{R}^9 \mapsto \boldsymbol{\hat{q}}_\mathcal{T}^\mathcal{M} \in \mathbb{R}^6$).
By applying rigid body decoupling strategies, the equivalent plant,  as seen by the feedback controller, is expected to be diagonal up until the desired bandwidth \citep{185Wal}. \\
\hspace*{2mm} For construction of a first-principle model-based feedback controller, a combination of a PID controller and a low-pass filter is considered for all six physical axes. Additionally, a standard rigid body feedforward is employed to the control system for further enhancement of position tracking performance. \\
\hspace*{2mm} Since the feedback controller, including the rigid body decoupling matrices, is based on the first-principles model, position tracking performance is limited by the neglected complex electro-mechanic phenomena which are present in the real system, but are very difficult to model from a first-principle view point.
In order to improve position tracking performance, the steady state solution of the residual dynamics is viewed as load disturbance, which is modelled using the GP framework. Consequently, the control policy to compensate for the neglected dynamics is estimated from experimental data using Gaussian regression, which allows for augmentation of the identified GP model in the currently employed rigid body feedforward controller as illustrated in Figure \ref{fig:CtrlLoop}.

\section{Gaussian-Process based learning of  disturbances }
\label{Section3}
\vspace*{-2mm}
\subsection{Disturbance characteristics and experiment design}
\label{sec:3:distCharact}
\begin{figure}[h]
	\begin{center}
		\includegraphics[width=0.75\textwidth,height=5.5cm]{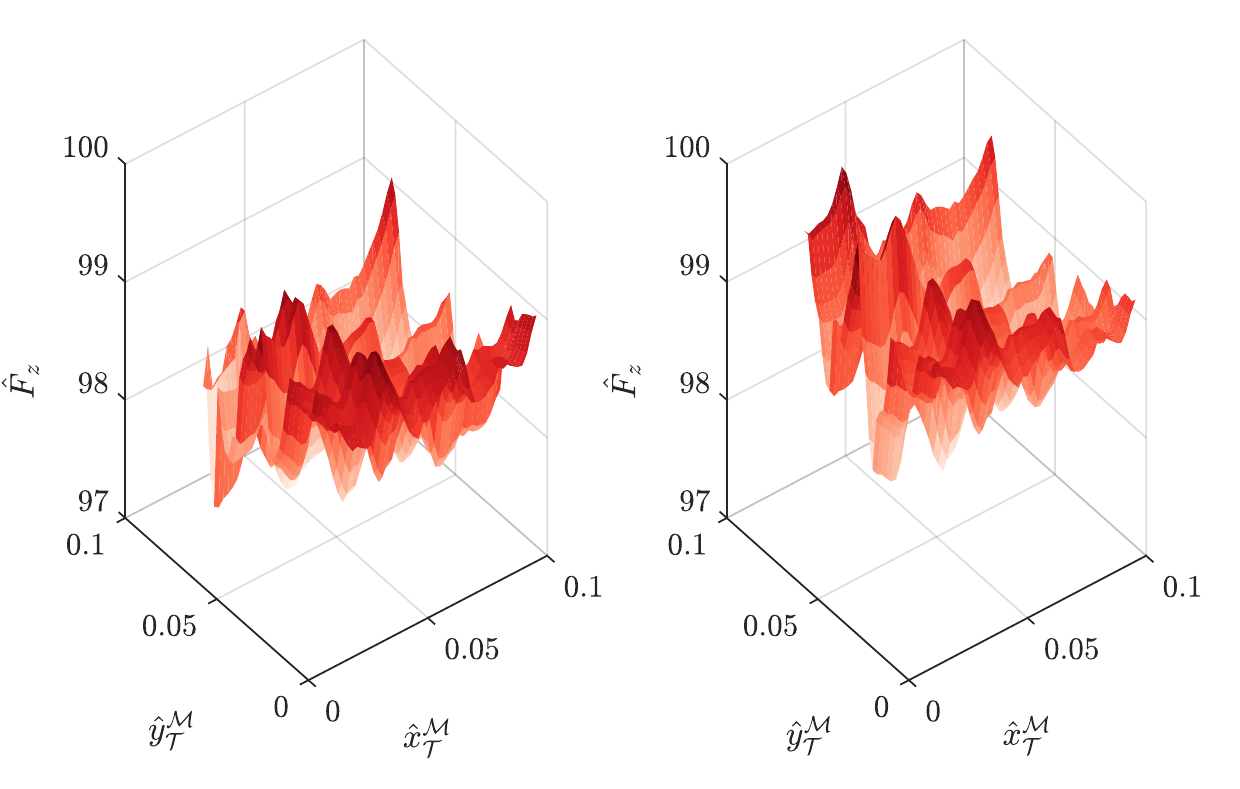}
		\caption{Two different measurements of $\hat{F}_{ z}$ on the $\hat{x}^{\mathcal{M}}_{\mathcal{T}}-\hat{y}^{\mathcal{M}}_{\mathcal{T}}$ plane.}
		\label{ch6:fig:fzGrid}
	\end{center}
\end{figure}

\vspace*{-1.5mm}
\hspace*{2mm} This section describes the modelling of static behaviour of residual dynamics using the GP framework. The main focus of this work is directed towards the suppression of disturbances acting on the $z$ axis, which exhibits the largest deviations from the desired position in the investigated planar motor due to effects of residual dynamics, see \citep{Proimadis-phd}. Apart from cancelling the known non-linear effects associated with the rigid-body dynamics, the commutation algorithm, presented in Section \ref{Section2}, further facilitated the control design by decoupling the system and reducing the number of inputs from $40$, i.e. the number of active coils, to $6$, i.e. the rigid-body of force and torque components. For similar reasons, it is desirable to view the static effect of residual dynamics acting on the $z$ axis as load disturbances acting on the output $\hat{F}_z$ of the RB decoupled feedback controller.
Using the Gaussian Process modelling framework, the disturbance force model is described by
\begin{equation}
\begin{aligned}
y(i)
=
{F}_{z}^{\mathrm{dist}}  \left( \boldsymbol{w} \left( i \right) \right)
+
e(i)
\end{aligned}
\label{eq:gpGenerateSystem},
\end{equation}
where $i$ is the time index and ${e}$ is the noise source, modelled as i.i.d. white Gaussian noise, ${e} (i) \sim \mathcal{N} \left( 0, \sigma_e^2 \right) $ \citep{DekkingBook}, while $y \in \mathbb{R}$ is the measured output. The disturbance force ${F}_{z}^{\mathrm{dist}}$ is assumed to be dependent on the input vector $\boldsymbol{w} \in \mathbb{R}^{n_u}$ and will be modelled by a Gaussian Process \citep{Rasmussen},
\begin{equation}
\begin{aligned}
{F}_{z}^{\mathrm{dist}} 
(\boldsymbol{w}) &\sim \mathcal{G}\mathcal{P} 
\big( 
0
,
k \left( \boldsymbol{w}(i), \boldsymbol{w}(j) \right)
\big),
\end{aligned}
\label{ch6:eq:gp_def2}
\end{equation}
i.e. its covariance is fully characterized by the kernel function $k$. The Gaussian Process framework exploits the assumption of Gaussian distribution of both the function to be estimated and the involved noise process, in order to arrive at an analytically computed predictor, which is based on the posterior predictive distribution of ${F}_{z}^{\mathrm{dist}}$ conditioned on  a training set of input-output observations. The mean of the predictive distribution is also the maximum a posteriori estimate
\begin{equation}
\bar{F}_{z}^{\mathrm{dist}} 
\left(\boldsymbol{w}^* \right) 
=
\boldsymbol{k}
\left(\boldsymbol{w}^*, W_N \right) 
\left( 
\mathcal{K} \left( W_N, W_N \right)
+ 
\sigma _e ^2 I_N \right)^{-1} \boldsymbol{y}, 
\end{equation}
where $\boldsymbol{w}^* \in \mathbb{R}^{n_u}$ is a test input, and $W_N$ with $\boldsymbol{y}$ contain the training input and output data, respectively, at the $N$ training points. For the vector $\boldsymbol{k}
\left(\boldsymbol{w}^*, W_N \right) \in \mathbb{R}^{1 \times N}$, its $i$\textsuperscript{th} element is equal to $k \left( \boldsymbol{w}^*, \boldsymbol{w} \left(i\right) \right)$, and in a similar fashion, for the matrix $\mathcal{K} \left( W_N, W_N \right) \in \mathbb{R}^{N \times N}$ the element on the  $i$\textsuperscript{th} row and $j$\textsuperscript{th} column is equal to $k \left( \boldsymbol{w} \left(i\right), \boldsymbol{w} \left( j \right) \right)$.
%
%
%
\\
\hspace*{2mm} Based on the discussion so far, three main questions have to be answered towards the modelling of the disturbance via the Gaussian Process framework. First, in spite of the model postulated in \eqref{eq:gpGenerateSystem}, the static effects of the residual dynamics cannot be directly measured. Consequently, we need a methodology to infer the non-measurable values of these static effects from measurable data. Secondly, the relevant inputs have to be defined. Thirdly, the kernel function has to be chosen.
\\
\hspace*{2mm} In order to answer the aforementioned questions, in \figref{ch6:fig:fzGrid}, the total force command on $z$ axis on a selected grid of positions in the $x^\mathcal{M}_\mathcal{T}-y^\mathcal{M}_\mathcal{T}$ plane is shown, after the dynamic behaviour has settled. The two plots correspond to two independent measurements. At steady state, due to the integral action in the feedback controller, the resulting total control effort in each DOF is equal to the feedforward effort, minus the constant disturbances that act on the same DOF, i.e.
\begin{equation}
\begin{aligned}
\hat{F}_{z} = \hat{F}_{z}^{\mathrm{FF}} - {F}_{z}^{\mathrm{dist}}
\end{aligned}
\label{eq:Force_effort},
\end{equation}
with $\ \hat{F}_{z}^{\mathrm{FF}}$ denoting the feedforward effort. By making use of \eqref{eq:Force_effort}, it becomes apparent that the measurements in \figref{ch6:fig:fzGrid} can be directly used to estimate the magnitude of the disturbance.
\newpage

\begin{figure}[h]
	\begin{center}
		\includegraphics[scale = 0.85]{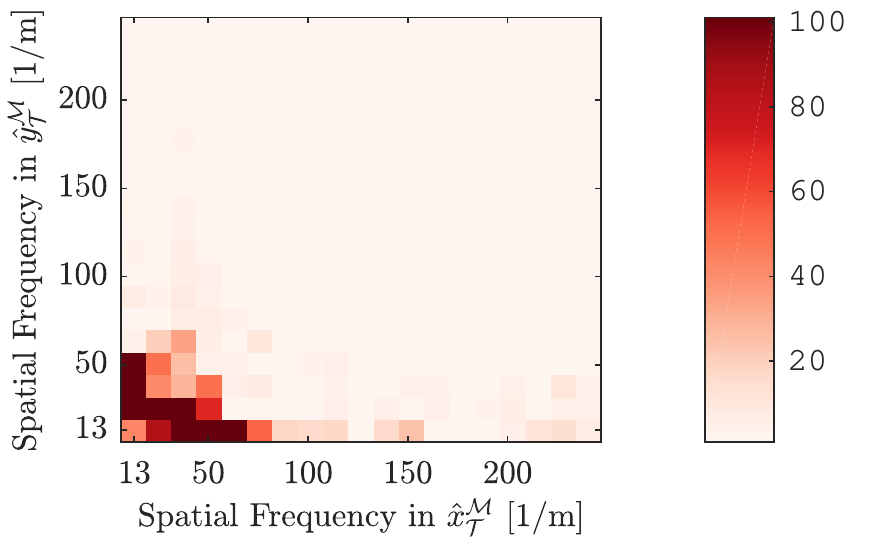}
		\caption{Spatial Fourier transform with respect to the $\hat{x}^{\mathcal{M}}_{\mathcal{T}},\hat{y}^{\mathcal{M}}_{\mathcal{T}}$ coordinates.}
		\label{ch6:fig:fft2DSpatial}
	\end{center}
\end{figure}

\vspace*{-7mm}
\noindent
\hspace*{2mm} \figref{ch6:fig:fzGrid} can be used further to draw conclusions, which are important for inferring which signals are affecting the static behaviour of residual dynamics, which are viewed as load disturbances. As a starting point, due to the position-dependent characteristics of the planar motor behaviour, it is natural to assume that the disturbances exhibit position-dependent characteristics as well. Secondly, it is asserted that the load disturbance exhibits a sinusoidal behaviour over the  $x^\mathcal{M}_\mathcal{T}-y^\mathcal{M}_\mathcal{T}$ plane. This is verified by computing the 2D Fourier transform of the measured $\hat{F}_{z}$ with respect to the latter coordinates, shown in \figref{ch6:fig:fft2DSpatial}, where it becomes evident that a sinusoidal behaviour with a spatial frequency of approximately $2.5-3$ cm is dominant. This value is related to the expected magnet pitch, see \cite{Custers-phd}. 
Thirdly, the differences in the two measurements, presented in \figref{ch6:fig:fzGrid}, reveal that the residual dynamics cannot be solely described as a function of $\hat{x}^{\mathcal{M}}_{\mathcal{T}},\hat{y}^{\mathcal{M}}_{\mathcal{T}}$. The deviation per grid point is on average equal to $0.46 \mathrm{N}$, which is approximately equal to $0.5 \%$ of the total mass feedforward control force $\hat{F}_{z}^{\mathrm{FF}}$. Equation \eqref{eq:rigidbodyEM} reveals a plausible source of residual dynamics; the vector $\hat{\boldsymbol{W}}_{\mathcal{M}}$ depends on the relative position between the coil, metrology and translator frame. Consequently, any unforeseen, neglected or inaccurately modelled dynamics will depend, to some extent, on the relative distance between the aforementioned frames. Consequently, at least six input variables are required in order to relate the coordinate frames to each other.
\vspace*{-7mm}

\noindent
\hspace*{2mm} The attenuation capabilities of the resulting feedforward control depend on how well the input space has been explored. The variables $\hat{\boldsymbol{q}}_{\mathcal{T}}^{\, \mathcal{M}}$ are actively controlled, therefore it is straightforward to define the excitation region with respect to them. The experiments have been performed on a $\hat{x}^{\mathcal{M}}_{\mathcal{T}}-\hat{y}^{\mathcal{M}}_{\mathcal{T}}$ grid, with $\hat{x}^{\mathcal{M}}_{\mathcal{T}}, \hat{y}^{\mathcal{M}}_{\mathcal{T}} \in \left[ 0.01, 0.1 \right]$ m. The position difference between two consecutive measurement points is approximately equal to $2$ mm. On the other hand, the $\hat{z}^{\mathcal{M}}_{\mathcal{T}}$ as well as the rotation of the translator with respect to the metrology frame are kept constant during operation. Regarding the $\boldsymbol{q}^{\mathcal{C}}_{\mathcal{M}}$ coordinates, the mounting of the metrology frame on passively controlled air mounts, as  explained in Section \ref{Section2}, means that the relative displacement between the coil and the metrology frame cannot be actively controlled. Hence, the input space exploration is limited to the variability of the experiments, which is affected by environmental or other, unforeseen physical phenomena. For this reason, the experiment on the aforementioned $\hat{x}^{\mathcal{M}}_{\mathcal{T}}, \hat{y}^{\mathcal{M}}_{\mathcal{T}}$ grid has been repeated six times and the corresponding $\hat{\boldsymbol{q}}^{\mathcal{C}}_{\mathcal{M}}$ have been estimated. In total, the input vector is defined as

\vspace*{-5mm}

\begin{equation}
\begin{aligned}
\boldsymbol{w} ( i)
=
\begin{bmatrix}
\hat{x}^{\mathcal{M}}_{\mathcal{T}}( i)
&
\hat{y}^{\mathcal{M}}_{\mathcal{T}}( i)
&
\hat{x}_{\mathcal{M}}^{\mathcal{C}}( i)
&
\hat{y}_{\mathcal{M}}^{\mathcal{C}}( i)
&
\hat{z}_{\mathcal{M}}^{\mathcal{C}}( i)
&
\hat{\chi}_{\mathcal{M}}^{\mathcal{C}}( i)
&
\hat{\psi}_{\mathcal{M}}^{\mathcal{C}}( i)
&
\hat{\zeta}_{\mathcal{M}}^{\mathcal{C}}( i)
\end{bmatrix}^{\top},
\end{aligned}
\label{ch6:eq:inputSignals}
\end{equation}
with $\boldsymbol{w} ( i) \in \mathbb{R}^8$ and $i = \left\{ 1, \ldots, N_p \right\}$, where $N_p$ denotes the number of data points. Moreover, ${w_\nu} ( i)$ denotes the $\nu^\text{th}$ element of the vector. Based on these inputs, an appropriate kernel selection is sought next.
\vspace*{-3mm}
\subsection{Kernel design and validation}
\vspace*{-1.5mm}
\hspace*{2mm}In the Gaussian Process framework, the kernel structure constitutes an important aspect, since it specifies the hypothesis space for the resulting estimator. As such, it is beneficial to incorporate any
observations to specify the structure of the expected function class. To this end, the aforementioned load disturbance characteristics can be inscribed in the kernel structure, thus facilitating the fitting of an appropriate GP model to the experimental data. Based on the experimental observations in Section~\ref{sec:3:distCharact}, the kernel is defined as the product of a periodic kernel \citep{384Mackay}, which captures the coil array characteristics, a Radial Basis Function (RBF) kernel \citep{NealBook}, which captures inconsistencies in the magnetic force distribution and a linear kernel, which explores linear trends with respect to the  $\hat{x}^{\mathcal{M}}_{\mathcal{T}}$- 
$\hat{y}^{\mathcal{M}}_{\mathcal{T}}$ frame, which would 
cause $\hat{z}^{\mathcal{M}}_{\mathcal{T}}$ to vary as a function of $\hat{x}^{\mathcal{M}}_{\mathcal{T}}$,
$\hat{y}^{\mathcal{M}}_{\mathcal{T}}$ . The resulting kernel is mathematically described by
\begin{equation}
\begin{aligned}
k 
\left( 
\boldsymbol{w} \left( i \right), 
\boldsymbol{w} \left( j \right) 
\right)
=&
\sigma_{1}^2 \exp 
\bigl(
-
\overbrace{
\sum\limits_{\nu = 1}^{8} 
\frac{
	\left( 
	{w_\nu} \left( i \right) 
	-
	{w_\nu} \left( j \right)
	\right)^2
}
{
	2 {\lambda}_{\nu, \mathrm{rbf}}^2
}}^{\text{RBF kernel}}
-
\overbrace{
\sum\limits_{\nu = 1}^{8} 
\frac{
	\sin^2
	\left(
	\frac{
		\pi
		\left(
		{w_\nu} \left( i \right) 
		-
		{w_\nu} \left( j \right)
		\right)
	}{
		p_{\mathrm{sin}}
	}
	\right)
}
{
	2 \lambda_{\mathrm{sin}}^2
}}^{\text{Periodic kernel}}
\bigl)
+
\\
& \underbrace{\sum\limits_{\nu = 1}^{2} 
{\sigma}_{\nu, 2}^2
\left(
{w_\nu} \left( i \right) 
-
{c}_{\nu, \mathrm{lin}}
\right)
\left(
{w_\nu} \left( j \right) 
-
{c}_{\nu, \mathrm{lin}}
\right)}_{\text{Linear kernel}},
\end{aligned}
\label{ch6:eq:kernelFun}
\end{equation}
where $\sigma_{\nu, 2}, {c}_{\nu, \mathrm{lin}}$ is the $\nu^{\text{th}}$ element of the vector $\boldsymbol{\sigma}_{2}, \boldsymbol{c}_{\mathrm{lin}}$ respectively. In total, there are 16 hyperparameters to be optimized, including the noise variance $\sigma_e$. Additionally, variations of the kernel structure, given by  (\ref{ch6:eq:kernelFun}) have been investigated. More specifically, the kernel structure in \eqref{ch6:eq:kernelFun} without the sinusoidal term, i.e. linear plus RBF, has been investigated, as well as a standalone RBF kernel. These two kernel structures are comprised of $14$ and $10$ hyperparameters, respectively. For all the aforementioned kernels, the hyperparameters have been computed by maximizing the marginal likelihood with respect to the hyperparameters \citep{Rasmussen}, using the collected training set.
\\
\hspace*{2mm} For the validation of the results, 5 data sets with sufficient variations in $\boldsymbol{q}^{\mathcal{C}}_{\mathcal{M}}$ are used for training purposes, where in total 3600 training points have been used. The predicted output $\hat{\boldsymbol{y}}$ is compared to the measured output $\boldsymbol{y}$ on a fresh data set using the Best Fit Ratio (BFR) 
\begin{equation}
\begin{aligned}
\text{BFR} &= 100 \% 
\cdot 
\max
\left(
1 
- 
\frac{
	\left| \left|
	\boldsymbol{y} - \hat{\boldsymbol{y}}
	\right| \right|_2
}{
	\left| \left|
	\boldsymbol{y} - \bar{{y}}
	\right| \right|_2}
,
0
\right),
\end{aligned}
\label{ch6:eq:bfr}
\end{equation}
where $\bar{y}$ denotes the sample mean of $\boldsymbol{y}$. The BFR criterion delivers an estimate between $0\%$ for no match and $100\%$  for the perfect match between the estimated and the measured output. The corresponding BFR results are presented in Table~\ref{tab:comparisonKernels}.
\begin{table}[h]
	\begin{center}
		\caption{Validation of the estimated GP compensator on the validation data set}
		\label{tab:comparisonKernels}
		\begin{tabular}{c|ccc}
			\hline
			{Kernel}                            
			& 
			{Linear + RBF $\times$ Sinusoidal}
			&
			{Linear + RBF}
			&
			{RBF}
			\\
			\hline \hline
			BFR $\%$
			& 
			85.77
			& 
			77.60
			&
			77.62
			\\
			\hline
		\end{tabular}
	\end{center}
\end{table}

\noindent
\hspace*{2mm} A few conclusions can be drawn from Table~\ref{tab:comparisonKernels}. First of all, it is evident that the employed kernels manage to efficiently capture the major characteristics of the disturbances that act on the motor prototype. Among the three kernels, the kernel containing the sinusoidal term leads to the highest accuracy, thus verifying the physics-based intuition for selecting such a kernel. However, the remaining two kernels also manage to predict the steady state behaviour of residual dynamics in a satisfactory manner, which highlights the flexibility of the RBF kernel in capturing non-linear trends. Finally, it is observed that any linear trend can be captured by the RBF kernel alone, thus explaining the similarity in results for the Linear+RBF and RBF kernels. In total, the results show that the GP framework is a powerful tool for modelling highly complex, position-dependent static effects of residual dynamics, which can severely limit the positioning accuracy of planar motors.

\section{Experimental results}
\begin{figure}[h!]
	\begin{center}
		\includegraphics[width=\textwidth,height=6cm]{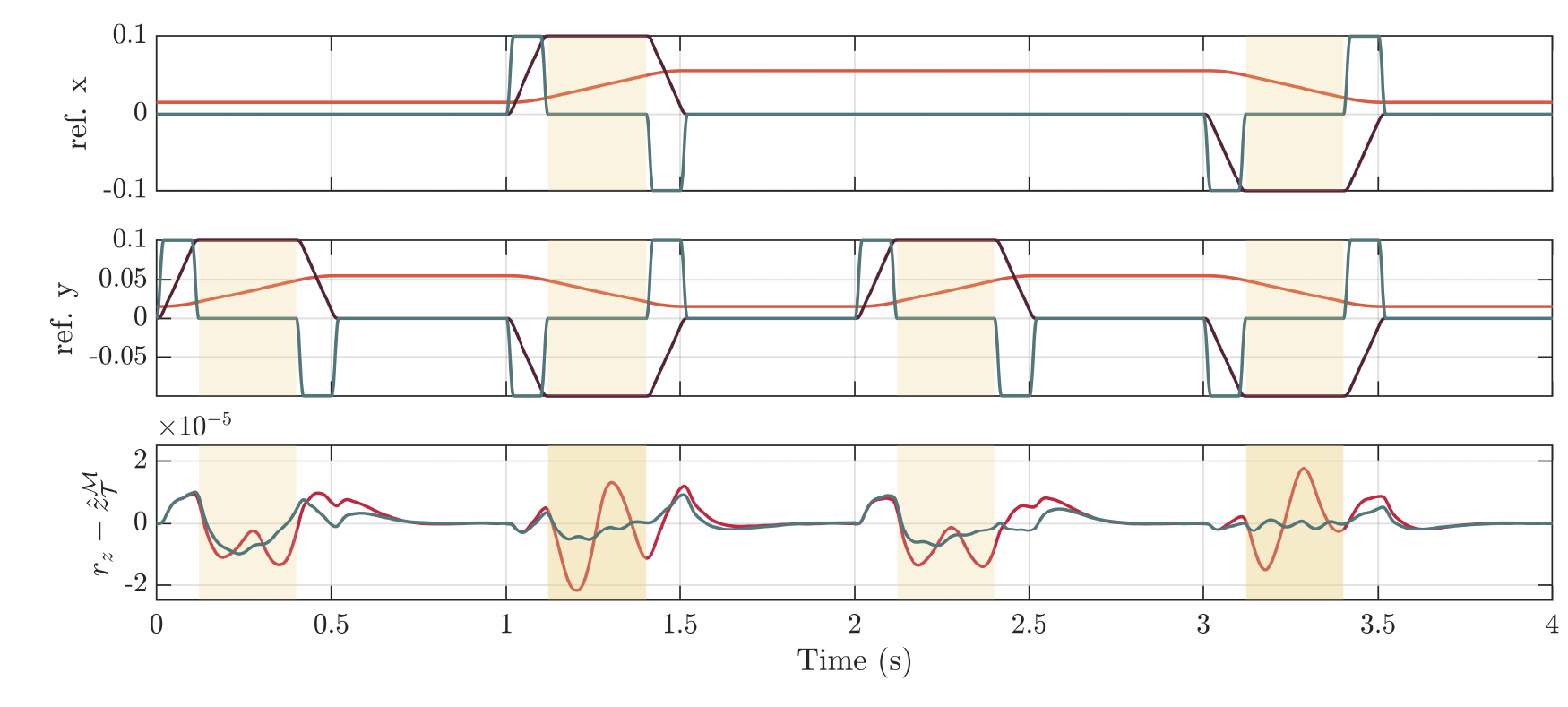}
		\caption{From top to bottom: Reference trajectory in terms of position $[m]$ (\myRule{thoughtProvRed}), velocity $[\frac{m}{s}]$ (\myRule{thoughtProvDark}) and scaled acceleration $[\frac{m}{s^2}]$ by a factor of 10 (\myRule{thoughtProvGreen}) for $\hat{x}^{\mathcal{M}}_{\mathcal{T}}$. 
			Reference position $[m]$ (\myRule{thoughtProvRed}), velocity $[\frac{m}{s}]$ (\myRule{thoughtProvDark}) and scaled acceleration $[\frac{m}{s^2}]$ by a factor of 10 (\myRule{thoughtProvGreen}) for $\hat{y}^{\mathcal{M}}_{\mathcal{T}}$. Error in $z$-direction $[m]$ with (\myRule{thoughtProvGreen}) and without (\myRule{thoughtProvRed}) kernel-based feedforward compensation for static disturbance rejection. The yellow areas represent the time intervals for which the velocity is constant.}
		\label{ch6:fig:MlStaticExpResults}
	\end{center}
\end{figure}

\vspace*{-2mm}
\hspace*{2mm} In this section, the GP-based predictor is used for compensation of  static disturbances, which are induced by residual dynamics that the feedback controller must compensate. Real-time implementation of the feedforward augmentation on the experimental prototype requires additional consideration. First of all, for $\hat{x}^{\mathcal{M}}_{\mathcal{T}}$, $\hat{y}^{\mathcal{M}}_{\mathcal{T}}$, the corresponding reference signals are used in the predictor instead of their measurement-based estimates. This selection is justified by the fact that for these two controlled variables, the error in this high-precision motor is relatively small compared to the sensitivity of the GP with respect to these variables. Moreover, introduction of additional feedback control loops is avoided, which could otherwise endanger stability. Additionally, the total control effort, shown in \figref{ch6:fig:fzGrid}, contains the feedforward control effort, i.e. the gravity compensation term. Therefore, in order to only compensate for the disturbance, the gravity compensation term is subtracted from the predicted disturbance force. 
\\
\hspace*{2mm} In order to evaluate the attenuation capabilities of the proposed feedforward controller  in the NAPAS prototype, its performance is tested under both diagonal as well as straight motion profiles with respect to the $\hat{x}^{\mathcal{M}}_{\mathcal{T}}$-$\hat{y}^{\mathcal{M}}_{\mathcal{T}}$ plane. 
For experimental validation of the designed feedforward augmentation, a low bandwidth feedback controller ($\approx$ 9.5 Hz) is applied to obtain a better visualization of the effects of the designed feedforward augmentation on the position tracking performance. The selected plane is defined as $\hat{x}^{\mathcal{M}}_{\mathcal{T}}$,$\hat{y}^{\mathcal{M}}_{\mathcal{T}} \in [0.015, 0.055] \mathrm{m}$. The employed 4\textsuperscript{th} order reference trajectory is generated based on the method by \cite{208Lambrechts}. The designed reference trajectory is illustrated in \figref{ch6:fig:MlStaticExpResults}. 
\\
\hspace*{2mm} For the real-time implementation of the GP-based feedforward augmentation, the computational complexity of the predictor has to be taken into consideration. For demonstration purposes, the linear plus RBF kernel is implemented in the experimental prototype, which, in practice, offers a good balance between prediction accuracy and required real-time calculation time. To further reduce the computation time, the initial predictor, which is formulated using the $3600$ training points, is further approximated using a subset of $200$ points. To accomplish this approximation, the Subset of Regressors (SR) \citep{143Wahba} method is applied, for which is empirically shown \cite[Chapter 8]{Rasmussen} that it offers a good balance between approximation error and computation time. For the selection of the subset of training points, a heuristic approach is considered \citep{Proimadis-phd}.
First, 200 samples are randomly selected for which the corresponding kernel is constructed. Secondly, the prediction capability of the resulting predictor is validated on a fresh data set by computing the BFR value. The heuristic approach is repeated 1000 times, after which the predictor with the highest BFR is kept. Using this approach, the approximated predictor achieved a BFR value of $65.37\%$, compared to the original BFR value of  $77.60\%$. 
\\
\hspace*{2mm} The experimental results are shown in Figure \ref{ch6:fig:MlStaticExpResults}. The trajectory consists of a motion on the $+{y}^{\mathcal{M}}_{\mathcal{T}}$ direction, followed by a diagonal motion on the $+{x}^{\mathcal{M}}_{\mathcal{T}}, -{y}^{\mathcal{M}}_{\mathcal{T}}$ direction, another $+{y}^{\mathcal{M}}_{\mathcal{T}}$ motion and finally the magnet plate returns to the starting point. Moreover, the error on the $\hat{z}^{\mathcal{M}}_{\mathcal{T}}$ axis is plotted for two cases, namely when the GP-based feedforward augmentation is active and inactive. Due to the coupling between the various DOFs, it is observed that the motion in the ${x}^{\mathcal{M}}_{\mathcal{T}}$ and/or ${y}^{\mathcal{M}}_{\mathcal{T}}$ axes leads to higher error in $\hat{z}^{\mathcal{M}}_{\mathcal{T}}$, too, compared to steady-state intervals. As a consequence, attenuating the disturbances in the former intervals is more imperative, especially in the constant velocity part, for which scanning takes place in the lithography process. Indeed, it can be directly observed in Figure \ref{ch6:fig:MlStaticExpResults} that the GP-based feedforward augmentation leads to significant improvement both in position tracking performance during steady-state, as well as position tracking performance during motion. 
\\
\hspace*{2mm} Finally, in order to quantify the resulting improvement with the GP-based feedforward augmentation, the $\ell_2$ norm, scaled by the square root of the number of data points, and the $\ell_\infty$ norms of the $\hat{z}^{\mathcal{M}}_{\mathcal{T}}$  error are computed. In Table~\ref{ch6:tab:comparisonNormStaticDistCompFfConstVel} the constant velocity results are presented, while in Table~\ref{ch6:tab:comparisonNormStaticDistCompFfWholeTraj} the results for the whole trajectory are presented. Based on the results shown in both tables, it is asserted that the improvement is approximately $50 \%$, thus highlighting the effectiveness of the proposed approach towards achieving higher positioning accuracy.

\begin{table}[h!]
	\begin{center}
		\caption{Experimental results of the feedforward augmentation.}
		\label{ch6:tab:comparisonNormStaticDistCompFfWholeTraj}
		\begin{tabular}{cccc}
			\hline
			&
			{No compensation}                            
			& 
			{Kernel-based FF}
			&
			Relative reduction
			\\
			\hline
			\hline
			{$\ell _2 / \sqrt{N}$ } 
			&
			$6.4 \cdot 10^{-6}$
			&
			$3.39 \cdot 10^{-6}$
			&
			$47.03 \%$
			\\
			{$\ell _\infty$}
			&
			$2.19  \cdot 10^{-5}$
			&
			$9.97 \cdot 10^{-6}$
			&
			$54.47 \%$
			\\
			\hline
		\end{tabular}
	\end{center}
\end{table}

\newpage

\vspace*{-7mm}
\begin{table}[h]
	\begin{center}
		\caption{Experimental results of the feedforward augmentation during constant velocity.}
		\label{ch6:tab:comparisonNormStaticDistCompFfConstVel}
		\begin{tabular}{cccc}
			\hline
			&
			{No compensation} 
			& 
			{Kernel-based FF}
			&
			Relative reduction
			\\
			\hline
			\hline
			{$\ell _2 / \sqrt{N}$ }                           
			&
			$1.02 \cdot 10^{-5}$
			&
			$4.38 \cdot 10^{-6}$
			&
			$57.06\%$
			\\
			{$\ell _\infty$}
			&
			$2.19  \cdot 10^{-5}$
			&
			$9.97 \cdot 10^{-6}$
			&
			$54.47\%$
			\\
			\hline
		\end{tabular}
	\end{center}
\end{table}
\label{ch:ExperimentalResults}

\section{Conclusions}
\vspace*{-3mm}
\hspace*{2mm} This paper presents a kernel based modelling approach that is able to successfully capture the steady state behaviour of residual dynamics for moving-magnet planar actuator systems as a function of the generalized coordinates associated to the coil - metrology and the translator - metrology reference frames. Additionally, capturing the static behaviour of residual dynamics using the GP framework allows for augmentation of the feedforward controller, such that unforeseen static effects of residual dynamics are timely attenuated for. Consequently, position tracking error in $z_\mathcal{T}^\mathcal{M}$ direction is reduced by more than 50 $\%$ during application.\\
\hspace*{2mm} Finally, in order to further improve position tracking performance of a planar motor stage, the kernel based modelling strategy can be extended, such that the full dynamic residual model is considered for feedforward compensation.
\label{ch:Conclusions}
\newpage

\bibliography{./MyBib}

\end{document}